%% file: main_icse.tex
\documentclass[10pt,conference]{IEEEtran}

\usepackage[nocompress]{cite}
\usepackage{algorithmic}
\usepackage{array,multirow,graphicx}
\usepackage{url}
\usepackage{amsthm, amsmath,amssymb,amsfonts}
\usepackage{textcomp}
\usepackage{xcolor}
\usepackage{xspace}
\usepackage{booktabs}
\usepackage{tabulary}
\usepackage{balance} 
\usepackage{threeparttable}
\usepackage{multirow}
\usepackage{hyperref}
\usepackage[utf8x]{inputenc}
\usepackage{enumitem}
\usepackage{ulem}
\usepackage{soul}
\usepackage{wasysym}
\usepackage{fdsymbol}
\usepackage{siunitx}
\usepackage{svg}
\usepackage{svg-extract}
\usepackage{subfig}
\usepackage{graphicx}
\usepackage{textcomp}
\theoremstyle{definition}
\newtheorem{definition}{Definition}[section]

\newcommand{\F}{$F_\beta$\xspace}

\def\BibTeX{{\rm B\kern-.05em{\sc i\kern-.025em b}\kern-.08em
    T\kern-.1667em\lower.7ex\hbox{E}\kern-.125emX}}
\begin{document}

\sisetup{
  table-number-alignment = center,
  table-figures-integer = 1,
  table-figures-decimal = 3
}

\title{Leveraging Transformer-based Language Models to Automate Requirements Satisfaction Assessment\\
\thanks{Identify applicable funding agency here. If none, delete this.}
}

\author{\IEEEauthorblockN{Amrit Poudel, Jinfeng Lin, Jane Cleland-Huang}
\textit{University of Notre Dame}\\
Notre Dame, IN \\
\{apoudel, jling, janeclelandhuang\}@nd.edu}

\maketitle
\begin{abstract}
Requirements Satisfaction Assessment (RSA) evaluates whether the set of  design elements linked to a single requirement provide sufficient coverage of that requirement -- typically meaning that all concepts in the requirement are addressed by at least one of the design elements. RSA is an important software engineering activity for systems with any form of hierarchical decomposition -- especially safety or mission critical ones. In previous studies, researchers used basic Information Retrieval (IR) models to decompose requirements and design elements into chunks, and then evaluated the extent to which chunks of design elements covered all chunks in the requirement. However, results had low accuracy because many critical concepts that extend across the entirety of the sentence were not well represented when the sentence was parsed into independent chunks. In this paper we leverage recent advances in natural language processing to deliver significantly more accurate results. We  propose two major architectures: Satisfaction BERT (Sat-BERT), and Dual-Satisfaction BERT (DSat-BERT), along with their multitask learning variants to improve satisfaction assessments. We perform RSA on five different datasets and compare results from our variants against the chunk-based legacy approach. All BERT-based models significantly outperformed the legacy baseline, and Sat-BERT delivered the best results returning an average improvement of 124.75\% in Mean Average Precision.
\end{abstract}


\input{01-Introduction.tex}
\input{02-ProblemDefinition.tex}

\input{03-ResearchQuestions.tex}

\input{04-Approach}
\input{05-ExperimentalEvaluation}
\input{06-ResultsDiscussion}

\input{07-RelatedWork}

\input{08-ThreatstoValidity}

\input{09-ConclusionandFutureWork}

\section{Data Availability}
  All of the datasets used in our work are either linked to their publicly available sources (i.e., CM1, Dronology, CHHIT or are proprietary and used under an NDA (PTC-A, PTC-B). We will release all source code for running experiments upon publication. Releasing it now during DBR is difficult due to the need to temporarily obfuscate all programmer information.

\bibliography{icse_bibs}
\bibliographystyle{IEEEtran}

\end{document}

%% file: 01-Introduction.tex
\section{Introduction}
\label{sec:intro}
\begin{figure}[t]
    \centering
    \includegraphics[width=.85\columnwidth]{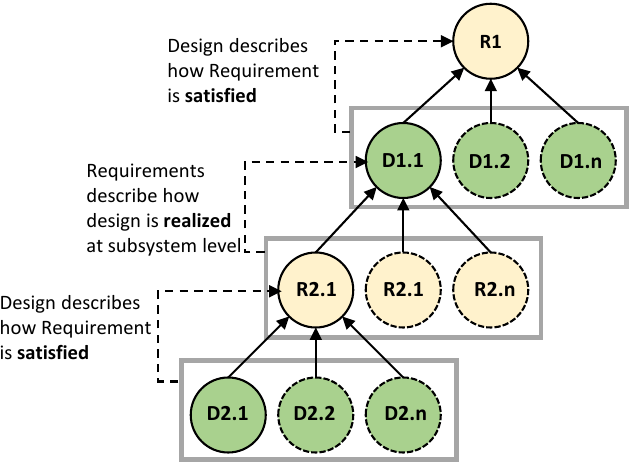}
    \caption{Requirements and design are refined into layers, in which lower level artifacts must fully refine (or satisfy) the higher level artifact to which they are linked.}
    \label{fig:refinement}
    \vspace{-12pt}
\end{figure}

Complex systems are typically refined into multiple layers of requirements and design \cite{DBLP:journals/software/WhalenGCMHR13, DBLP:conf/re/WhalenMH12}, such that artifacts at one level in the hierarchy are refined or decomposed by lower level artifacts at the next level. As depicted in Figure \ref{fig:refinement}, the layering of requirements and design continues down to the leaf nodes at which point design elements are realized through lower level models or implemented code. In order to demonstrate that top-level artifacts have been satisfactorily addressed, the lower-level artifacts in each layer are linked back to the higher-level artifacts from which they are derived. These trace links are typically stored in a document known as the Requirements Traceability Matrix (RTM) \cite{DBLP:conf/icse/Cleland-HuangGHMZ14,DBLP:books/daglib/p/GotelCHZEGDAMM12}. While trace links can be created manually, they require significant cost and effort to create and maintain; therefore much research effort has been vested in automating the process through the use of information retrieval (e.g., \cite{DBLP:conf/iwpc/OlivetoGPL20,DBLP:journals/tse/HayesDS06} and more recently, deep-learning techniques \cite{guo2017semantically,lin2021traceability,lin2022enhancing}. The majority of prior work on automated tracing techniques has focused on the primitive task of generating a link between pairs of related artifacts -- for example, determining whether a given design artifact should be linked to a specific requirement. Very little research effort has focused on the more challenging traceability-related task of RSA which determines whether a requirement has been fully satisfied by its set of lower-level artifacts.

The process of determining whether a set of natural language design element/s linked to a natural language requirement provides sufficient coverage to satisfy the requirement is termed Requirements Satisfaction Assessment (RSA) \cite{holbrook2006assessing}. From an automation perspective, we consider a requirement to be satisfied when all the major semantic concepts that it contains are present in the set of design elements. 

While satisfaction assessment has remained relatively understudied for the last decade, it is of great importance in safety, or mission critical systems \cite{holbrook2009toward} where identifying requirements that are not fully satisfied by their related design elements can reduce the need for costly redesign efforts, and help to ensure that all identified hazards have been satisfactorily addressed. Organizations therefore often hire analysts or even external evaluators to validate and verify that all specified requirements are fully addressed by their respective design elements \cite{cleland2014software}. This activity is often supported through the creation of a Requirements Trace Matrix, which maps requirements to design elements via trace links \cite{gotel1996extended}. The trace matrix is then used by safety analysts and external certifiers to help assess whether requirements, especially safety-related ones, are satisfied in the designed and deployed system. The fact that the process is time-consuming and error-prone makes the use of automation appealing.

\subsection{Legacy work in RSA automation}
In prior work, Holbrook et al. 
\cite{holbrook2009toward} explored several NLP approaches for automating the RSA process. They deconstructed both the requirement and its linked design artifacts into chunks, computed the similarity between design chunks and requirement chunks, and then checked the extent to which each requirement chunk was covered by one or more of the design chunks. They explored both a simple term-based approach for matching chunks as well as a thesaurus-enhanced approach \cite{holbrook2009toward}. They also proposed a rule-based approach that used parts-of-speech (POS) tagging to detect lexico-syntactic patterns \cite{holbrook2013study}; however, this type of approach  relies upon specific grammatical structures in the requirements and design text, and requires non-trivial user effort to prepare the domain-specific project thesaurus and to establish grammar rules for matching chunks. 

In general, techniques that rely heavily on chunking-based approaches, n\"aively compare chunks as isolated elements, and therefore fail to capture the full semantics of each artifact. As a result, they fail to bridge the semantic gap that exists between different types of artifacts \cite{guo2017semantically,lin2021traceability} and subsequently do not deliver accurate results.

\subsection{Proposed Solution} 
Fortunately, recent advances in NLP have provided effective techniques for representing sentences as a whole while  preserving their underlying syntactic and semantic information. One particularly effective technique, leverages pretrained language models (PLMs) to represent each sentence in its entirety in a latent vector space.
 PLMs are large neural networks \cite{elazar2021measuring} typically trained using self-supervised learning on a huge corpus of data. They are often applied within a two-phase paradigm (i.e, \emph{pretrain-finetune}) that includes pretraining followed by a finetuning step in which model weights are first trained on a huge corpus of data, and then later optimized for a specific task (e.g., RSA) through a \emph{downstream training task} that is closely related to the targeted application. In this way, relevant knowledge is transferred from the pretrained model to the downstream domain-specific model. This paradigm has been used extensively to address diverse NLP tasks such as Question Answering, Machine Translation, and Text Entailment \cite{devlin2018bert, liu-etal-2019-xqa}. Furthermore, the use of PLMs in conjunction with the \emph{pretrain-finetune} paradigm has been shown to be effective for automating the basic tracing task in which links are generated between pairs of artifacts \cite{lin2021traceability}.  To the best of our knowledge, these techniques have not yet been applied to improve the accuracy of automated RSA.
 
In this paper, we use PLMs within several different RSA solutions to identify cases of missing or superfluous design elements. We also replicate several of the IR techniques from Holbrooke's work in order to create a legacy baseline, and then comparatively evaluate automated RSA solutions based on two BERT architectures (Vanilla-BERT and Dual-BERT) and the multitasking variant (MSat-BERT). 
Our experiments are conducted against artifacts from five datasets of hierarchically decomposed requirements, and results show that all BERT models far outperform Holbrook's legacy solution, and that our multi-task learning solution provides significant further enhancements. 

The major contributions of this paper are therefore as follows:
\begin{itemize}
\item We reconstruct a legacy baseline based on ideas from Holbrook's prior work on RSA.
\item We propose several NLP approaches using pretrained language models and multitask learning (MLT) to improve the accuracy of RSA.
\item We conduct a series of experiments against five datasets and demonstrate very significant improvements over prior IR-based approaches. Results from our experiments lay down a new baseline, and refocus attention on a greatly understudied problem.
\end{itemize}

The remainder of the paper is structured as follows. In Section \ref{sec:Problem}, we formally introduce the problem of RSA and define several research questions. In Section \ref{sec:Approach}, we describe our techniques, and then evaluate them experimentally in Section \ref{sec:ExperimentalEvaluation}.  Section \ref{sec:ResultsDiscussions} discusses results, and Sections \ref{sec:RelatedWork} to \ref{sec:Conclusion} discuss related work, threats to validity, and finally draw conclusions. 

%% file: 02-ProblemDefinition.tex
\section{RSA Problem Definition}
\label{sec:Problem}
In this section we describe the RSA task through providing concrete examples and identifying specific RSA challenges.

\subsection{Motivating Examples}
We start by illustrating the RSA process with two examples, that serve to indicate how challenging RSA is to automate. Our first example is taken from the the autonomous pilot of a small Unmanned Aerial System (sUAS). Requirement R1 has been linked to three different design artifacts labeled D1-D3 as follows:
\begin{itemize} [leftmargin=6.5mm]
    \item[\bf R1] When the sUAS receives a go\_to\_waypoint command, it flies to the targeted GPS coordinates at the specified velocity.
    \item[\bf D1] Upon receipt of a waypoint command, the sUAS plans a trajectory for its flight.
    \item[\bf D2] The sUAS executes the flight plan at the defined speed.
    \item[\bf D3] When the computed distance from the sUAS to the target waypoint is less than {{THRESHOLD\_WAYPOINT\_DISTANCE}} than the waypoint is considered reached. 
\end{itemize}

All three design elements explain how the waypoint command will be executed and include information about planning the trajectory, setting the velocity, and determining when the waypoint is reached. The problem we observe here is that all three design elements are relevant; however, based purely on semantic analysis,  a non-domain expert might find D2 sufficient for satisfying R1 if D1 and D3 were missing.

Our second example, is taken from the Isolette System \cite{Isolette}, where system level requirement {\it SR} has been linked to three design artifacts {\it (D5-D7)}. For illustrative purposes we also add design artifact {\it (D8)}.

\begin{itemize} [leftmargin=6.5mm]
    \item[\bf R2] The Thermostat must be self-tested at startup by pressing the self-test button.
    \item[\bf D5] The system starts in the INIT mode when it is powered on and remains in this mode until the Regulator Status becomes valid.
    \item[\bf D6] The regulator must self-test during initialization.
    \item[\bf D7] If the system does not complete its initialization within a specified time-out period it enters FAILED mode.
    \item[\bf D8] The alarm sounds when the isolette exceeds maximum\_temperatures.
\end{itemize}

A preliminary analysis of the requirement shows that it describes the need for self-tests at startup. D1 is related to SR because it describes the startup process (i.e., {\it powered on}) and the fact that it waits for regulator status to become valid. D2 is clearly related as it specifically refers to the {\it self-test}, while D3 is related because it also talks about the initialization process. However, there are two problems.  First, D8 does not seem to be related, and furthermore, the `red button' referenced in R2 is not addressed in any of the design elements.  We conclude that design elements (D5-D7) are correctly linked to the requirement, D8 is a misfit, and further that the requirement is not fully satisfied through the set of design elements.

\subsection{Challenges}
Based on these motivating examples, we formally introduce two specific RSA automation challenges (C1 and C2), and a third hybrid challenge (C3). The problems associated with each challenge typically occur when a project stakeholder either fails to create appropriate trace links during the development process or fails to fully consider how to realize the requirement in the design.

\begin{figure}[t]
\vspace{-6pt}
    \centering
   \includegraphics[width=.85\columnwidth]{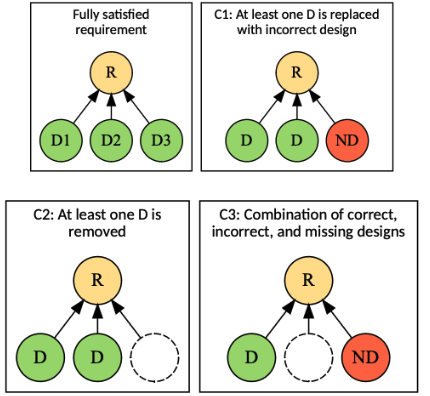}
   \caption{Three different experimental challenges (C1, C2, C3) where R = Requirement, D = Correctly linked design element, ND = incorrectly linked design element, and dashed circle = missing design element.}
   \label{fig:challenges}

\end{figure}

\subsection*{C1: Replaced Design Element}
This scenario is illustrated in Fig \ref{fig:challenges}.C1 and occurs when a correct link is replaced by an incorrect one.  The problem is defined more formally as follows:
\begin{definition}
\textit{Given a requirement R, and its corresponding set of linked design elements, where at least one of the related design elements D has been replaced by an unrelated design element ND s.t. R = \{D, D, ND\}. The automated RSA model is responsible for classifying whether requirement R is satisfied or unsatisfied.}
\end{definition}

\subsection*{C2: Missing Design Element}
This scenario is illustrated in Fig. \ref{fig:challenges}.C2, and represents the case where one or more design elements is missing. This is a common scenario, as project stakeholders may neglect to add links, or may fail to fully decompose the requirement into a complete set of design artifacts. We formally define the challenge as follows:
\begin{definition}
\textit{Given a requirement R, and its corresponding set of incomplete design elements where at least one design element is missing s.t. R = \{D, D, $\phi$\}; the satisfaction model is used to classify requirement R as satisfied or unsatisfied.}
\end{definition}

\subsection*{C3: Hybrid Scenario}
Finally, we explore a hybrid scenario in which design artifacts may be missing and/or replaced by unrelated design elements. We formulate C3 as a multi-class classification problem, in order to predict whether the set of design elements is corrupt, incomplete, or complete, where ``incomplete' implies at least one missing design element, ``corrupt'' means that an unrelated element is included, and ``complete'' implies that the set of design artifacts satisfy the requirements. We formally define the problem as follows:
\begin{definition}
\textit{Given a requirement R, and its corresponding set of design elements where either one of the design elements has been replaced by an incorrect design element or is missing or removed, s.t. R = \{D, D, ND\} or R = \{D, D, $\phi$\} respectively; the satisfaction model classifies the requirement R as corrupt, complete, or incomplete.}
\end{definition}

%% file: 03-ResearchQuestions.tex
\subsection{Research Questions}
\label{sec:RQ}

In this study, we propose several different techniques based on the use of pretrained language models for solving these three RSA-related challenges. We comparatively evaluate these models through addressing the following research questions:
    
\noindent\textbf{RQ1: To what extent does the use of a pretrained language model outperform the legacy approach?}

We then explore the use of dual-BERT encoders, in which additional information is fed to the model to see if different forms of information help the model to learn how to perform RSA better. Our research question is formulated as follows:
\noindent\textbf{RQ2:How does the dual-BERT architecture perform in comparison to the single-BERT architecture?} 

All of these approaches assume single-task learning for the fine-tuning phase; however, Caruana et al., \cite{caruana1997multitask} argued that single-task learning potentially fails to fully utilize all of the rich information provided in the input string. We therefore explore multi-task learning by incorporating an additional, closely-related task of link prediction to the primary task of requirements satisfaction analysis. We address the following research question:
\noindent\textbf{RQ3:  Does Multitask Learning (MTL) return more accurate results in comparison to single-task learning?} 

Finally, after conducting a detailed analysis of all our proposed techniques we determine which model is the best overall model for solving the general RSA challenge across all of our datasets through the following research question:
\noindent\textbf{RQ4: Which BERT-based model returns the best results for the general RSA challenge?}

%% file: 04-Approach.tex
\section{Approach}
\label{sec:Approach}
Before describing our individual techniques we briefly summarize the fundamental concepts of the BERT architecture. We then present our three solution models, namely: Sat-BERT, MSat-BERT, and DSat-BERT.

\subsection{Language Models and BERT}
Language models are designed to predict the next elements of a sequence. In particular, given a sentence with a sequence of words \{w$_1$, w$_2$, ..., w$_k$\}, a Language Model (LM) calculates the probability of the next word in the sequence i.e. P(w$_k$ $\mid$ $w_{1:k-1}$) or the probability of the whole sequence i.e. P(w$_{1:k}$). LMs can therefore be considered as a probability distribution over a sequence of words\cite{jurafskyspeech}. 

Transformers are attention-based mechanisms \cite{vaswani2017attention} that have substantially improved a number of diverse natural language processing (NLP) tasks. For example, Devlin et al. \cite{devlin2018bert} proposed an encoder-based language representation model called Bidirectional Encoders Representations from Transformers, widely known as BERT. BERT is trained on two tasks. The first is Masked Language Modeling (MLM) in which word(s) are randomly masked and the model is trained to replace them with the same words, and the second is Next Sentence Prediction (NSP) in which the model is trained to understand dependencies across sentences thereby alleviating the problem of unidirectionality. 

BERT is trained using the pretrain-finetune paradigm. In the pretraining stage, the BERT-based model is fed with a huge corpus of data, and then fine-tuned using labeled data related to the targeted downstream tasks. To create a unified architecture for performing different tasks, BERT can be fine-tuned by adding an additional output layer commonly referred to as the \textbf{task head} \cite{devlin2018bert}.

\subsection{BERT for Requirements Satisfaction Assessment}
For purposes of automating the RSA problem we explore three different BERT architectures i.e. Sat-BERT, DSat-BERT, MSat-BERT for RSA. In this naming scheme, `Sat' stands for Requirements SATisfaction analysis, and therefore Sat-BERT and DSat-BERT represent vanilla solutions using single and dual architectures applied to the RSA problem. In contrast, MSat-BERT is an extension that incorporates multi-tasking. For each of these architectures we experiment with three general purpose language models - BERT \cite{devlin2018bert}, sciBERT \cite{beltagy2019scibert} and seBERT \cite{lin2022enhancing}. While these models share the same underlying principles, they have been trained using different training data to support domain-specific tasks. BERT is trained on a general corpus of data i.e. Wikipedia and BookCorpus \cite{zhu2015aligning}, while sciBERT is trained on a large multi-domain corpus of scientific publications, and seBERT on a general software engineering domain obtained from millions of OSS projects on GitHub. In the following sub-sections, we discuss the architectures in details.

\begin{figure}[h]
  \begin{center}
    \includegraphics[width=.65\columnwidth]{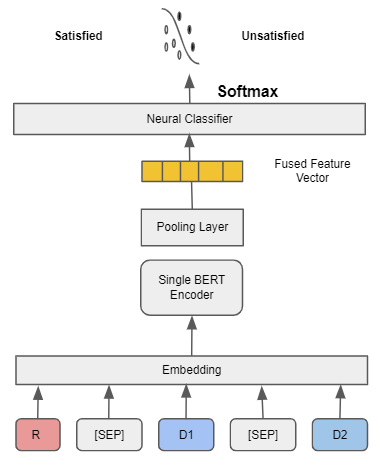}
  \end{center}
  \caption{Sat-BERT}
  \label{fig:satBERT}
\end{figure}

\begin{figure}[h]
  \begin{center}
  \includegraphics[width=.875\columnwidth]{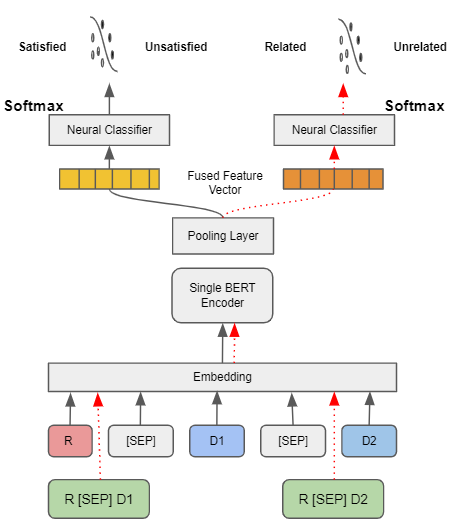}
  \end{center}
  \caption{MSat-BERT}
  \label{fig:msatBERT}
\end{figure}

\begin{figure}[h]
  \begin{center}
    \includegraphics[width=1\columnwidth]{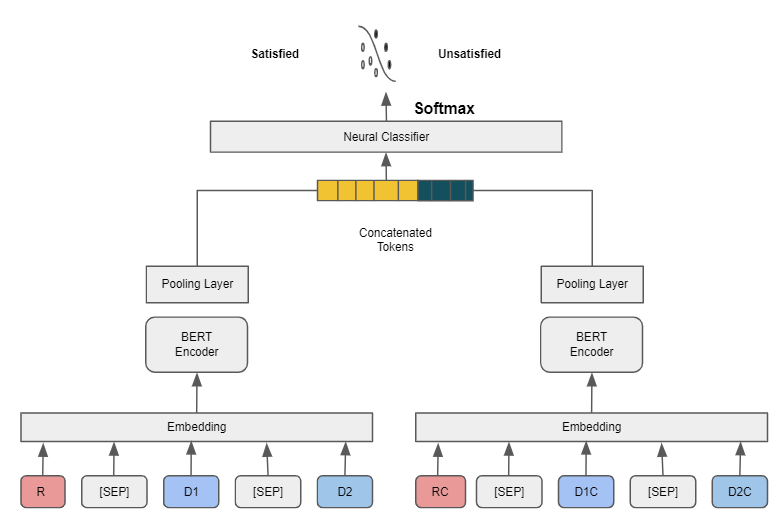}
  \end{center}
  \caption{DSat-BERT}
  \label{fig:dsatBERT}
\end{figure}


\subsubsection{\textbf{Sat-BERT}}
As depicted in Fig \ref{fig:satBERT}, Sat\_BERT utilizes a single BERT model to encode inputs. All the design elements linked to a corresponding requirement are concatenated with the requirement they are linked to as follows. First, the tokens of the design elements $D_1$ \& $D_2$ are concatenated with a special token [SEP] between each element. Second, the resulting string is concatenated to the requirement tokens forming a single sequence $I_{sat}$ s.t. $I_{sat}$ = $[CLS]$ \textit{$r_1$, $r_2$, ..., $r_N$} $[SEP]$ \textit{$d_{1,1}$, $d_{1,2}$, ..., $d_{1,N}$} $[SEP]$ \textit{$d_{2,1}$, $d_{2,2}$, ..., $d_{2,N}$} $[SEP]$, where $[CLS]$ and $[SEP]$ are special tokens marking the start and the end of the sentences respectively. 

The transformed representation of the requirement-design pair i.e. $I_{sat}$  is then fed to the BERT model, which generates a hidden matrix of 768 dimensions (default for BERT). The hidden matrix is passed through the series of feed-forward neural networks, which compresses the matrix into a lower dimension. Finally, a classification head categorizes the input into the specified number of classes by assigning the probabilities calculated using the Softmax function to produce a probability distribution. 

\subsubsection{\textbf{MSat-BERT}}
MSat-BERT is an extension to the Sat-BERT, in which the auxiliary task of link prediction is added by including an extra classification head in the single BERT model as shown in Fig \ref{fig:msatBERT}. The use of auxiliary tasks introduces multitask learning. The rationale for utilizing multitasking for finetuning a model is based on the way humans learn. Intuitively, humans first learn to solve easy problems, and then use the knowledge they have acquired to solve progressively more complex problem\cite{ruder2017overview}. 
Caruana et al., \cite{caruana1997multitask} described the use of related skills to master a primary task as an inductive transfer mechanism, and demonstrated better generalization across various tasks using Multitask Learning (MTL). They argued that many aspects of rich information sources are not leveraged in single-task Learning, whereas MTL, utilizes the information contained in training signals from different tasks as long as they are drawn from the same domain. There are two different ways to perform MTL, referred to as hard-parameter and soft-parameter sharing. Hard-parameter sharing involves sharing the hidden layers among various tasks, while having distinct task-specific layers, whereas soft-parameter sharing involves regularizing the distance between the task-specific model parameters in order to encourage the parameters to be similar \cite{ruder2017overview}. 
We adopt the most commonly used approach i.e. hard-parameter sharing as it improves generalization, and has a low risk of overfitting since the tasks are learned simultaneously \cite{baxter1997bayesian}. Therefore, each of our two tasks,  i.e. satisfaction prediction \textit{(main)} and trace link prediction \textit{(auxiliary)} are assigned their own layers.  \vspace{5pt}

\textbf{MSat-BERT} follows the same transformation as Sat-BERT for the main task i.e. satisfaction prediction; however, for the link prediction, instead of concatenating all design elements to the requirement they are linked to, we perform a pairwise combination of each requirement to each of its design elements s.t. $I_{link}$ = \{$[CLS]$ \textit{$r_1$, $r_2$, ..., $r_N$} $[SEP]$ \textit{$d_{1,1}$, $d_{1,2}$, ..., $d_{1,N}$} $[SEP]$, $[CLS]$ \textit{$r_1$, $r_2$, ..., $r_N$} $[SEP]$ \textit{$d_{2,1}$, $d_{2,2}$, ..., $d_{2,N}$} $[SEP]$\}. After the transformation, the shared BERT model is fed with both $I_{sat}$ and $I_{link}$ which outputs two respective hidden state matrices. The matrices are then passed to their respective task-specific classification head, and the individual losses are obtained via Cross Entropy loss. Mathematically, 

If $L_{sat}$ \& $L_{link}$ are the two losses for satisfaction and link prediction tasks respectively, we minimize the multitask objective as:
\begin{equation}
    \begin{aligned}
    L = L_{sat} + \lambda L_{link} \\
    \end{aligned}
    \label{eq:multiloss}
\end{equation}

where $L_{sat}$, and $L_{link}$ are obtained via Cross Entropy (CE) Loss over their size of their respective training data. Mathematically,
        

\begin{equation}
    \begin{aligned}
        \text{CE} = -\sum_{n=1}^{\text{output size}} y_i \log \hat{y}_i
    \end{aligned}
    \label{eq:CE}
\end{equation}

In equation \ref{eq:multiloss}, we assign greater importance to the satisfaction task as compared to the auxiliary link prediction task, and therefore set the value of $\lambda $ to 0.5 for minimizing the multitask objectives in all the MTL related experiments.\\

\subsubsection{\textbf{DSat-BERT}}
As shown in Fig \ref{fig:dsatBERT}, the DSat-BERT architecture leverages two BERT models (M1,M2). In this proposed architecture, M1 is fed the unaltered natural language requirements and design elements as input, while M2 is fed the requirement and design chunks. We hypothesize that utilizing both types of input data will provide better generalizability, with each model paying attention to different types of concept. 

DSat-BERT implements a similar transformation technique to that of Sat-BERT. While the maximum sequence length for model M1 is set to the default size of 512, we shrink M2's maximum sequence length to 128, given that chunks are shorter in length. The outputs from the models M1 and M2 of size 512*768 and 128*768 respectively are then concatenated to form a hidden representation of the size 740*768. The concatenated matrix then goes through the same steps as that of Sat-BERT. \\

%% file: 05-ExperimentalEvaluation.tex
\section{Experimental Evaluation}
\label{sec:ExperimentalEvaluation}

\subsection{Datasets}

We evaluated our proposed architectures against five datasets  as shown in Table \ref{dataset}. We established inclusion criteria for selecting datasets as follows. First, as  RSA inherently assesses refinement and/or decomposition relationships, we only included datasets if they had a clear hierarchical structure. Second, data needed to generally exhibit 1:N relationships between the higher level artifact (e.g., requirement) and the lower-level artifact (e.g., design). This is expected but not guaranteed in a hierarchical structure. Third, we sought datasets with existing trace matrices that could serve as an answer set for our experimental evaluation, and fourth we favored datasets that provided coverage of different domains, in order to evaluate generalizability. 
\input{tables/data}

Our first dataset, \textbf{CM1}, consists of requirement and design elements for a NASA scientific instrument, and is publicly available at the Promise Software Engineering Repository \cite{Sayyad-Shirabad+Menzies:2005}. The \textbf{CCHIT} dataset, available from COEST.org provides requirements for digitizing patients records and information, and has previously been used in research relating to tracing regulatory codes \cite{DBLP:conf/icse/Cleland-HuangGHMZ14}. The \textbf{Dronology} \cite{cleland2018dronology} dataset includes requirements and design artifacts for 
managing and coordinating the flights of small Unmanned Aerial Systems (sUAS). It has been used in several previous research projects \cite{rahimi2018evolving, krismayer2019constraint, krismayer2019supporting}. We use only requirements and design definitions.  The final dataset is from a \textbf{Positive Train Control (PTC)} system, and was provided by our industrial collaborators under a non-disclosure agreement. It contains high level SRS (System requirements), mid level SDD  (system design), and  low level SSRS (subsystem requirements). Our PTC-A dataset uses links from SRS to SDD, while PTC-B uses links from SDD to SSRS.

\input{tables/RQ1}

\subsection{Experiment Setup}
We illustrate our experimental setup for a generalized project containing requirements, design elements, and the mapping between the requirements and the design elements i.e. RTM are given s.t. $R = \{R_1, R_2 .... R_n\}, \ DS = \{D_1, D_2, ......D_m\} \ \& \ RTM = \{R_1 = \{D_1, D_2, D_3\}, R_2 = \{D_6, D_{10}\},....\} $ 
For the remainder of this discussion we refer to the set of artifacts consisting of one requirement (R) and several Design elements as an \emph{RD-set}.

\input{tables/splits}

For scenario C1, we replace one or more design elements in the RD-set with elements randomly selected from the pool of unrelated design elements. The goal is for the RSA model to identify the RD-set as corrupt.
We create k negative samples for each requirement such that each requirement has k unique corrupted RD-set instances. We experimented with three values of \textit{k} i.e. \textbf{100}, \textbf{300}, and \textbf{500}, and cap the negative samplings at the maximum possible number of combinations if k combinations cannot be formed. More intuitively, for a requirement R2 which maps to the set \{$D_6$, $D_{10}$\} when k is set to 100, we generate 100 negative samples by replacing each of the design elements in an equal proportion i.e. each of the elements would be replaced 50 times. Some of the negative samples that can be generated using this strategy for R2 are:  \{$D_6$, $D_1$\},  \{$D_6$, $D_2$\},  \{$D_1$, $D_{10}$\},  \{$D_5$, $D_{10}$\}, and so on. 

For scenario C2, we remove one or more of the design elements from the RD-Set to create the negative set. The goal is for our RSA models to identify which RD-sets are incomplete.  We create negative samples for requirements linked to more than one design element by replacing all the combinations from one to (n-1) design elements, where n refers to the length of the design set. For a requirement R1, which has three design elements, we would be able to generate six negative samples i.e. \{$D_1$, $D_2$\}, \{$D_1$, $D_3$\}, \{$D_2$, $D_3$\}, \{$D_1$\}, \{$D_2$\}, \& \{$D_3$\}. We ignore cases which contain only one design element, as removing one element would create an empty set. 

Finally, we introduce the hybrid scenario, which contains a mix of RD-Sets representing correct, incorrect, and missing design elements. The goal is to differentiate between RD-Sets that are complete, corrupted, or incomplete. We create k number of negative samples using the negative sampling strategy from C1. In addition, we also create another batch of negative samples for each requirement by removing each unique design elements from the set with replacement. Finally, we concatenate the two batches of negative samples, therefore creating the C3 scenario. Intuitively, for a given mapping R1, the following negative samples can be generated on removing one unique design element with replacement: \{$D_1$, $D_2$\}, \{$D_2$, $D_3$\}, \{$D_1$, $D_3$\}. These generated samples are then mixed with the k negative samples generated using the sampling strategy from C1. 

For each scenario, after generating the dataset, we use the 'split-by-requirements' \cite{lin2022enhancing} approach for C1 and C3, where we randomly divide the requirements into ten folds (8:1:1) - 8 for training, and 1 each for test and validation set. We keep all RD-Sets related to a single requirement in the same fold. However, for C2, in order to compare results across equally sized RD-Sets, we use the 'split-by-dlength' where we first group the requirements by their respective number of design elements and then split them individually into ten folds (8:1:1). After the completion of the splits, we merge the bucketed folds from individual splits to form the train, test and validation set. Table \ref{splits} reports the size of the splits for each of the datasets and for each task.

To increase reliability of our results, we repeat this process five times using different seeds. Each experiment is repeated on each of the five unique splits, and we report the average scores from these runs. 


\subsection{Legacy Baseline}
In order to compare against an existing baseline, we closely replicated the chunking approach developed by Holbrook et al. \cite{holbrook2006assessing} with slight modifications.  We broke requirements and design elements into noun phrase chunks using the NLTK \cite{bird2004nltk} parser, lemmatized the chunks, and applied the pairwise cosine similarity between the requirement and design chunks. We categorized each requirement chunk as covered if it had a similarity score of above 0.5 with at least one of the design elements, and uncovered otherwise. We then calculated the overall coverage rate as follows:

\begin{equation}
    Coverage\ Rate = \frac{Total\ Covered\ Chunks}{Total\ Requirement\ Chunks}
    \label{eq:CoverageRate}
\end{equation}

Finally we computed Inverse Document Frequency (IDF) scores to calculate the importance of a term in a given document. IDF was calculated as the logarithm of the total number of documents to the total number of documents containing the term, meaning that rare terms receive higher IDF scores than common ones.   Following Holbrook et al. we treated each chunk as an individual document for calculating the idf scores \cite{holbrook2006assessing}.

In addition, we adopted Shapley values, as a concept from the field of game theory, where features are weighed according to their contribution to the given task \cite{wiki:Shapley_value}. We applied the Softmax function to assign importance to the requirement chunks based on their idf scores. Mathematically for a requirement R with its corresponding chunks $\{r_1, r_2, .., r_n\}$, first we calculate the idf scores for all the chunks, and then calculate the Softmax scores, thus converting the idf scores into normalized probabilities.
\begin{equation}
    Softmax = \frac{\exp{x_i}}{\sum_{j=1}^{n}\exp{x_j}}
    \label{eq:Softmax}
\end{equation}
We extract Softmax scores for the covered chunks obtained via cosine similarity, and calculate the coverage score as follows: 
\begin{equation}
    Coverage\ Score = \exp{\sum_{j}^{|C|}Softmax(x_j)}
    \label{eq:CoverageScore}
\end{equation}
In equation \ref{eq:CoverageScore},  $|C|$ refers to the cardinality of the set containing all the covered chunks. We show in our experiments that Coverage Score generalizes better than the n\"aive Coverage Rate in most of the cases. We again repeat our experiments five times holding the test data while concatenating training and validation data for calculating the idf scores.  We provide raw data, showing average scores across all five runs in an online document at https://tinyurl.com/RSA-Results. 

\input{tables/RQ2}

\subsection{Model Training}
We trained all  variants of our proposed models for 1 epoch with a batch size of 4 and gradient accumulation steps of 8. We set the initial learning rate to 5E-05 and applied a linear scheduler at run time. We trained all of our models on each of the five different training datasets, while monitoring the performance on their respective validation dataset. We used the best model obtained through these steps as evaluated using the corresponding test dataset, and compared the results across the proposed architectures. 


\subsection{Evaluation Metrics}
We report the F2-scores for the binary classification problems i.e. C1 and C2, and Mean Average Precision (MAP) for the multi-class classification problem (C3) for which no clear threshold value exists. 

\subsubsection{Mean Average Precision (MAP)}
The RSA models assign scores for each RD-set according to the likelihood that it is either missing an element or corrupt. We use these scores to generate a ranked list, from which Average Precision (AP) is computed based on the position of a satisfied RD-set (ground-truth). MAP is then computed as the mean of AP values across all RD-sets.  Mathematically,

\begin{equation}
    mAP = 1/n \sum_{k=1}^{k=n} AP_k
    \label{eq:map}
\end{equation}

where, n is the number of satisfied RD-set and $AP_k$ is the Average Precision of $k^{th}$ RD-sets.

\subsubsection{F-scores}
F2-scores represent the harmonic mean of Recall and Precision, whilst favoring recall over precision as this tends to be more important for software engineering tasks such as RSA \cite{DBLP:conf/icse/ShinHC15}.  We report F2-scores for challenges C1 and C2, but report the macro-averaged F-1 score i.e. the unweighted mean of all the F-1 scores per classes for the multi-class classification problem. Mathematically, 

\begin{equation}
    F_\beta = \dfrac{(1+\beta^2).precision.recall}{\beta^2.precision.recall}
    \label{eq:fbeta}
\end{equation}

\input{tables/RQ3}

%% file: tables/data.tex
\begin{table}[tbh!]
\addtolength{\tabcolsep}{-3.6pt}
	\centering
    \caption{Software Engineering projects used in our experiments. All of these datasets are hierarchical in nature, follow one-to-many relationships, and provide a manually curated Requirements Traceability Matrix (RTM) necessary for conducting RSA.} 
    \begin{tabular}{lllllll}

    \toprule \hline
        Project  & Description & Source & Count& Target&Count&Links\\ \hline
        CM1& Scientific instrument &SRS&155&SDS&150&155 \\
        CCHIT  & Electronic record system &Regul.& 47&SRS&139&47\\
        Dronology& Multi-UAV flight coord. &SRS&94&SDS &210&94\\ 
        PTC-A   & Subway signalling system & SSRS & 400 &SDD &320  &400 \\ 
        PTC-B & Subway signalling system & SDD & 501 &SRS &159 &501\\
         
        \hline
        \bottomrule
    \end{tabular}
    \label{dataset}
\end{table}

%% file: tables/RQ1.tex
\begin{table*}[t!]
\centering
\caption{Comparison of performance between various language models leveraging vanilla BERT architecture i.e. Sat-BERT on all challenges against their respective legacy ($\lozenge$) baselines. For C2, we report the $F_\beta$ and MAP scores for the requirements with four design elements, wherever applicable.}
\begin{tabular}{|l|l|cc|cc|cc|cc|cc|cc|cc|}
\hline
\multicolumn{2}{|c|}{\multirow{2}{*}{Challenges}} & \multicolumn{2}{c|}{CM1} & \multicolumn{2}{c|}{CCHIT} & \multicolumn{2}{c|}{Dronology} & \multicolumn{2}{c|}{PTC-A} & \multicolumn{2}{c|}{PTC-B} & \multicolumn{2}{c|}{Average Score} & \multicolumn{2}{c|}{Average Improve} \\

\multicolumn{2}{|c|}{} & \multicolumn{1}{c}{$F_\beta$} & \multicolumn{1}{c|}{MAP} & \multicolumn{1}{c}{$F_\beta$} & \multicolumn{1}{c|}{MAP} & \multicolumn{1}{c}{$F_\beta$} & \multicolumn{1}{c|}{MAP} & \multicolumn{1}{c}{$F_\beta$} & \multicolumn{1}{c|}{MAP} & \multicolumn{1}{c}{$F_\beta$} & \multicolumn{1}{c|}{MAP} & \multicolumn{1}{c}{$F_\beta$} & \multicolumn{1}{c|}{MAP} &
\multicolumn{1}{c}{$F_\beta$} & \multicolumn{1}{c|}{MAP} \\ 

\hline
\multirow{3}{*}{\rotatebox[origin=c]{90}{\parbox[c]{1.8cm}{\centering C1}}}
& legacy $\lozenge$ & 0.073 & 0.177 & 0.036 & 0.010 & 0.061 & 0.253 & 0.091 & 0.245 & 0.123 & 0.221 & 0.077 & 0.181 & - & - \\

& bert & 0.163 & 0.206 & 0.175 & 0.105 & 0.423 & 0.535 & 0.473 & 0.534 & 0.484 & 0.579 & 0.344 & 0.392 & 346.88\% & 115.93\% \\

& scibert & 0.255 & 0.263 & 0.262 & 0.287 & 0.473 & 0.612 & 0.509 & 0.591 & 0.524 & 0.599 & 0.405 & 0.470 & \textbf{426.07}\% & \textbf{159.35}\% \\

& sebert & 0.252 & 0.278 & 0.161 & 0.101 & 0.541 & 0.672 & 0.456 & 0.560 & 0.419 & 0.506 & 0.366 & 0.423 & 375.59\% & 133.31\% \\
\hline
\multirow{3}{*}{\rotatebox[origin=c]{90}{\parbox[c]{1.8cm}{\centering C2}}}
& legacy $\lozenge$ & 0.409 & 0.137 & 0.553 & 0.213 & 0.327 & 0.090 & 0.562 & 0.237 & 0.583 & 0.250 & 0.487 & 0.185 & - & - \\

& bert & 0.774 & 0.563 & 0.779 & 0.557 & 0.967 & 0.900 & 0.858 & 0.650 & 0.721 & 0.545 & 0.820 & 0.643 & 68.41\% & 246.84\% \\

& scibert & 0.983 & 0.950 & 0.900 & 0.833 & 1.000 & 1.000 & 0.892 & 0.750 & 0.733 & 0.552 & 0.902 & 0.817 &\textbf{ 85.21}\% &\textbf{ 340.77}\% \\

& sebert & 0.963 & 0.925 & 0.814 & 0.687 & 0.817 & 0.625 & 0.933 & 0.800 & 0.720 & 0.515 & 0.849 & 0.710 & 74.45\% & 283.24\% \\
\hline
\multirow{3}{*}{\rotatebox[origin=c]{90}{\parbox[c]{1.8cm}{\centering C3}}}
 & legacy $\lozenge$ & - & 0.178 & - & 0.010 & - & 0.252 & - & 0.246 & - & 0.177 & - & 0.173 & - & - \\

& bert & 0.377 & 0.222 & 0.355 & 0.129 & 0.443 & 0.405 & 0.436 & 0.465 & 0.491 & 0.551 & 0.420 & 0.354 & - & 105.24\% \\

& scibert & 0.398 & 0.268 & 0.366 & 0.136 & 0.452 & 0.503 & 0.446 & 0.478 & 0.466 & 0.556 & 0.426 & 0.388 & - & \textbf{124.75}\% \\

& sebert & 0.379 & 0.217 & 0.294 & 0.087 & 0.471 & 0.460 & 0.414 & 0.457 & 0.425 & 0.499 & 0.396 & 0.344 & - & 99.23\% \\
\hline
\end{tabular}
\label{icse:RQ1}
\end{table*}

%% file: tables/splits.tex
\begin{table}
\centering
\caption{Split of requirements into three folds- train, valid and test sets for each datasets. C1 and C3 follows the split-by-requirements strategy, whereas C2 follows split-by-dlength.}
\begin{tabular}{c|c c c| c c c}
\hline
          & \multicolumn{3}{c|}{C1/C3} & \multicolumn{3}{c}{C2}  \\ 

 & train & valid & test       & train & valid & test     \\ 
\hline
CM1       & 124   & 15    & 16         & 70    & 10    & 9        \\ 

CCHIT     & 37    & 5     & 5          & 17    & 9     & 8        \\ 

Dronology & 75    & 9     & 10         & 39    & 6     & 6        \\ 

PTC-A      & 320   & 40    & 40         & 87    & 12    & 12       \\ 

PTC1-B     & 400   & 50    & 51         & 53    & 6     & 8        \\
\hline
\end{tabular}
\label{splits}
\end{table}

%% file: tables/RQ2.tex

\begin{table*}[t!]
\centering
\caption{Comparison of performance between single and dual BERT architectures i.e. Sat-BERT and DSat-BERT (suffixed \textit{\_dual}) on all challenges with their respective vanilla BERT as the baselines ($\blacklozenge$).}

\begin{tabular}{|l|l|cc|cc|cc|cc|cc|cc|cc|}
\hline
\multicolumn{2}{|c|}{\multirow{2}{*}{Challenges}} & \multicolumn{2}{c|}{CM1} & \multicolumn{2}{c|}{CCHIT} & \multicolumn{2}{c|}{Dronology} & \multicolumn{2}{c|}{PTC-A} & \multicolumn{2}{c|}{PTC-B} & \multicolumn{2}{c|}{Average Score} & \multicolumn{2}{c|}{Average Improve} \\

\multicolumn{2}{|c|}{} & \multicolumn{1}{c}{$F_\beta$} & \multicolumn{1}{c|}{MAP} & \multicolumn{1}{c}{$F_\beta$} & \multicolumn{1}{c|}{MAP} & \multicolumn{1}{c}{$F_\beta$} & \multicolumn{1}{c|}{MAP} & \multicolumn{1}{c}{$F_\beta$} & \multicolumn{1}{c|}{MAP} & \multicolumn{1}{c}{$F_\beta$} & \multicolumn{1}{c|}{MAP} & \multicolumn{1}{c}{$F_\beta$} & \multicolumn{1}{c|}{MAP} &
\multicolumn{1}{c}{$F_\beta$} & \multicolumn{1}{c|}{MAP} \\ 
\hline
\multirow{6}{*}{\rotatebox[origin=c]{90}{\parbox[c]{1.8cm}{\centering C1}}}
& bert $\blacklozenge$ & 0.163 & 0.206 & 0.175 & 0.105 & 0.423 & 0.535 & 0.473 & 0.534 & 0.484 & 0.579 & 0.344 & 0.392 & \multicolumn{1}{l}{-} & \multicolumn{1}{l|}{-} \\

& bert\_dual & 0.185 & 0.202 & 0.221 & 0.184 & 0.450 & 0.587 & 0.433 & 0.534 & 0.456 & 0.570 & 0.349 & 0.416 & 1.52\% & 6.08\% \\

& scibert & 0.255 & 0.263 & 0.262 & 0.287 & 0.473 & 0.612 & 0.509 & 0.591 & 0.524 & 0.599 & 0.405 & 0.470 & \textbf{17.72\%} & \textbf{20.11\%}\\

& scibert\_dual & 0.249 & 0.279 & 0.225 & 0.271 & 0.440 & 0.530 & 0.487 & 0.552 & 0.510 & 0.585 & 0.382 & 0.444 & 11.20\% & 13.22\% \\

& sebert & 0.252 & 0.278 & 0.161 & 0.101 & 0.541 & 0.672 & 0.456 & 0.560 & 0.419 & 0.506 & 0.366 & 0.423 & 6.42\% & 8.05\% \\

& sebert\_dual & 0.240 & 0.266 & 0.184 & 0.122 & 0.510 & 0.596 & 0.400 & 0.526 & 0.369 & 0.452 & 0.340 & 0.393 & -0.95\% & 0.22\% \\

\hline
\multirow{6}{*}{\rotatebox[origin=c]{90}{\parbox[c]{1.8cm}{\centering C2}}}
& bert $\blacklozenge$ & 0.774 & 0.563 & 0.779 & 0.557 & 0.967 & 0.900 & 0.858 & 0.650 & 0.721 & 0.545 & 0.820 & 0.643 & \multicolumn{1}{l}{-} & \multicolumn{1}{l|}{-} \\

& bert\_dual & 0.759 & 0.593 & 0.933 & 0.800 & 0.858 & 0.650 & 0.854 & 0.707 & 0.802 & 0.672 & 0.841 & 0.684 & 2.61\% & 6.47\% \\

& scibert & 0.983 & 0.950 & 0.900 & 0.833 & 1.000 & 1.000 & 0.892 & 0.750 & 0.733 & 0.552 & 0.902 & 0.817 & 9.98\% & \textbf{27.08\%} \\

& scibert\_dual & 0.938 & 0.833 & 1.000 & 1.000 & 0.789 & 0.580 & 0.824 & 0.629 & 0.909 & 0.767 & 0.892 & 0.762 & 8.82\% & 18.49\% \\

& sebert & 0.963 & 0.925 & 0.814 & 0.687 & 0.817 & 0.625 & 0.933 & 0.800 & 0.720 & 0.515 & 0.849 & 0.710 & 3.59\% & 10.50\% \\

& sebert\_dual & 0.938 & 0.833 & 1.000 & 1.000 & 0.854 & 0.707 & 0.876 & 0.667 & 0.909 & 0.767 & 0.915 & 0.795 & \textbf{11.67\%} & 23.61\% \\

\hline
\multirow{6}{*}{\rotatebox[origin=c]{90}{\parbox[c]{1.8cm}{\centering C3}}}
& bert $\blacklozenge$ & 0.377 & 0.222 & 0.355 & 0.129 & 0.443 & 0.405 & 0.436 & 0.465 & 0.491 & 0.551 & 0.420 & 0.354 & \multicolumn{1}{l}{-} & \multicolumn{1}{l|}{-} \\

& bert\_dual & 0.375 & 0.213 & 0.358 & 0.101 & 0.448 & 0.397 & 0.426 & 0.459 & 0.438 & 0.495 & 0.409 & 0.333 & -2.76\% & -6.04\% \\

& scibert & 0.398 & 0.268 & 0.366 & 0.136 & 0.452 & 0.503 & 0.446 & 0.478 & 0.466 & 0.556 & 0.426 & 0.388 & \textbf{1.21\%} & \textbf{9.51\%} \\

& scibert\_dual & 0.395 & 0.237 & 0.361 & 0.176 & 0.474 & 0.502 & 0.424 & 0.466 & 0.444 & 0.512 & 0.420 & 0.378 & -0.16\% & 6.81\% \\

& sebert & 0.379 & 0.217 & 0.294 & 0.087 & 0.471 & 0.460 & 0.414 & 0.457 & 0.425 & 0.499 & 0.396 & 0.344 & -5.72\% & -2.93\% \\

& sebert\_dual & 0.383 & 0.238 & 0.324 & 0.102 & 0.457 & 0.439 & 0.382 & 0.418 & 0.386 & 0.449 & 0.386 & 0.329 & -8.13\% & -7.10\% \\
\hline
\end{tabular}
\label{icse:RQ2}
\end{table*}

%% file: tables/RQ3.tex
\begin{table*}[t!]
\centering
\caption{Comparison of performance between vanilla BERT and MTL BERT architectures i.e. Sat-BERT and MSat-BERT (suffixed \textit{\_mul}) on all challenges with their respective vanilla BERT as the baselines ($\blacklozenge$).}

\begin{tabular}{|l|l|cc|cc|cc|cc|cc|cc|cc|}
\hline
\multicolumn{2}{|c|}{\multirow{2}{*}{Challenges}} & \multicolumn{2}{c|}{CM1} & \multicolumn{2}{c|}{CCHIT} & \multicolumn{2}{c|}{Dronology} & \multicolumn{2}{c|}{PTC-A} & \multicolumn{2}{c|}{PTC-B} & \multicolumn{2}{c|}{Average Score} & \multicolumn{2}{c|}{Average Improve} \\

\multicolumn{2}{|c|}{} & \multicolumn{1}{c}{$F_\beta$} & \multicolumn{1}{c|}{MAP} & \multicolumn{1}{c}{$F_\beta$} & \multicolumn{1}{c|}{MAP} & \multicolumn{1}{c}{$F_\beta$} & \multicolumn{1}{c|}{MAP} & \multicolumn{1}{c}{$F_\beta$} & \multicolumn{1}{c|}{MAP} & \multicolumn{1}{c}{$F_\beta$} & \multicolumn{1}{c|}{MAP} & \multicolumn{1}{c}{$F_\beta$} & \multicolumn{1}{c|}{MAP} &
\multicolumn{1}{c}{$F_\beta$} & \multicolumn{1}{c|}{MAP} \\ 
\hline
\multirow{6}{*}{\rotatebox[origin=c]{90}{\parbox[c]{1.8cm}{\centering C1}}}
& bert $\blacklozenge$ & 0.163 & 0.206 & 0.175 & 0.105 & 0.423 & 0.535 & 0.473 & 0.534 & 0.484 & 0.579 & 0.344 & 0.392 & - & - \\

& bert\_multi & 0.217 & 0.272 & 0.233 & 0.179 & 0.442 & 0.566 & 0.462 & 0.557 & 0.493 & 0.564 & 0.369 & 0.428 & 7.44\% & 9.18\% \\

& scibert & 0.255 & 0.263 & 0.262 & 0.287 & 0.473 & 0.612 & 0.509 & 0.591 & 0.524 & 0.599 & 0.405 & 0.470 & \textbf{17.72\%} & \textbf{20.11\% }\\

& scibert\_multi & 0.274 & 0.288 & 0.270 & 0.224 & 0.478 & 0.636 & 0.499 & 0.573 & 0.481 & 0.542 & 0.400 & 0.453 & 16.51\% & 15.58\% \\

& sebert & 0.252 & 0.278 & 0.161 & 0.101 & 0.541 & 0.672 & 0.456 & 0.560 & 0.419 & 0.506 & 0.366 & 0.423 & 6.42\% & 8.05\% \\

& sebert\_multi & 0.298 & 0.359 & 0.172 & 0.109 & 0.520 & 0.626 & 0.470 & 0.555 & 0.402 & 0.469 & 0.373 & 0.424 & 8.41\% & 8.14\% \\
\hline
\multirow{6}{*}{\rotatebox[origin=c]{90}{\parbox[c]{1.8cm}{\centering C2}}}

& bert $\blacklozenge$ & 0.774 & 0.563 & 0.779 & 0.557 & 0.967 & 0.900 & 0.858 & 0.650 & 0.721 & 0.545 & 0.820 & 0.643 & - & - \\

& bert\_multi & 0.908 & 0.813 & 0.714 & 0.630 & 1.000 & 1.000 & 0.850 & 0.700 & 0.906 & 0.813 & 0.876 & 0.791 & 6.83\% & 23.03\% \\

& scibert & 0.983 & 0.950 & 0.900 & 0.833 & 1.000 & 1.000 & 0.892 & 0.750 & 0.733 & 0.552 & 0.902 & 0.817 & 9.98\% & 27.08\% \\

& scibert\_multi & 0.943 & 0.867 & 1.000 & 1.000 & 0.911 & 0.840 & 0.967 & 0.900 & 0.809 & 0.600 & 0.926 & 0.841 & \textbf{12.95\%} & \textbf{30.87\%} \\

& sebert & 0.963 & 0.925 & 0.814 & 0.687 & 0.817 & 0.625 & 0.933 & 0.800 & 0.720 & 0.515 & 0.849 & 0.710 & 3.59\% & 10.50\% \\

& sebert\_multi & 0.929 & 0.864 & 0.816 & 0.767 & 1.000 & 1.000 & 0.858 & 0.650 & 0.670 & 0.385 & 0.855 & 0.733 & 4.24\% & 14.06\% \\
\hline

\multirow{6}{*}{\rotatebox[origin=c]{90}{\parbox[c]{1.8cm}{\centering C3}}}
& bert $\blacklozenge$ & 0.377 & 0.222 & 0.355 & 0.129 & 0.443 & 0.405 & 0.436 & 0.465 & 0.491 & 0.551 & 0.420 & 0.354 & - & - \\

& bert\_multi & 0.369 & 0.249 & 0.352 & 0.098 & 0.431 & 0.437 & 0.420 & 0.442 & 0.431 & 0.469 & 0.401 & 0.339 & -4.73\% & -4.30\% \\

& scibert & 0.398 & 0.268 & 0.366 & 0.136 & 0.452 & 0.503 & 0.446 & 0.478 & 0.466 & 0.556 & 0.426 & 0.388 & \textbf{1.21\%} & \textbf{9.51\%} \\

& scibert\_multi & 0.370 & 0.273 & 0.346 & 0.130 & 0.432 & 0.516 & 0.423 & 0.450 & 0.440 & 0.510 & 0.402 & 0.376 & -4.33\% & 6.00\% \\

& sebert & 0.379 & 0.217 & 0.294 & 0.087 & 0.471 & 0.460 & 0.414 & 0.457 & 0.425 & 0.499 & 0.396 & 0.344 & -5.72\% & -2.93\% \\

& sebert\_multi & 0.387 & 0.257 & 0.321 & 0.091 & 0.421 & 0.464 & 0.398 & 0.423 & 0.404 & 0.466 & 0.386 & 0.340 & -8.12\% & -3.99\% \\
\hline
\end{tabular}
\label{icse:RQ3}
\end{table*}

%% file: 06-ResultsDiscussion.tex
\section{Results and Discussions}
\label{sec:ResultsDiscussions}
We now address the  research questions from Section \ref{sec:Approach} based on our analysis of the results. 

\noindent\textbf{RQ1: To what extent does the use of pretrained language models outperform the legacy approach based on Information Retrieval?} 

We evaluated our techniques using three general purpose LMs - `BERT-base-uncased', `sciBERT-uncased', and `seBERT' - against five different datasets as listed on Table ~\ref{dataset}. For the C1 and C3 challenges, we report the average of MAP and F scores from three samples i.e. 100, 300, and 500 as our MAP and $F_\beta$ scores, whereas for the C2 challenge, we report the MAP and $F_\beta$ scores of requirements having four design elements (maximum among all datasets) in Table \ref{icse:RQ1}, \ref{icse:RQ2}, and \ref{icse:RQ3}. Complete results for C2 are available in our supplemental materials (\url{https://tinyurl.com/RSA-Results}). The column entitled ``Average Score'' calculates the average score across all five datasets, whilst ``Average Improve'' reports the percentage gain as compared to the legacy baseline. 

Among all the general purpose models, sciBERT achieved the greatest average improvements i.e. 426.07\% ($F_\beta$) and 159.35\% (MAP)  for C1, 85.21\% ($F_\beta$), and 340.77\% (MAP) for C2, and 124.75\% (MAP) for C3 over the legacy baseline. The results are reported in Table \ref{icse:RQ1}. This is likely because sciBERT is pretrained on multi-domain scientific corpus, which matches the domain of all five of our datasets. However, for Dronology, seBERT also performed well, likely due to the fact that it is trained on millions of open-source github projects, and there are far larger numbers of open-source projects related to drones, than to more obscure domains such as PTC and CM1.
In conclusion, results clearly show that pretrained LMs outperformed the IR-based approach in solving RSA challenges, and that overall sciBERT performed the best.

\noindent\textbf{RQ2:How does the dual-BERT architecture perform in comparison to the single-BERT architecture?} 
In this section, we address the second research question by comparing the performance of dual-BERT architecture with that of single-BERT. Results are  reported in Table \ref{icse:RQ2}.

We observed that in most cases the dual variants performed worse than their vanilla variant. For the C1 challenge, vanilla sciBERT performed better than all other variants, and improved over vanilla BERT by 17.72\% (\F) and 20.11\% (MAP), while its dual counterpart only achieved a gain of 11.20\% (\F) and 13.22\% (MAP). Vanilla sciBERT also outperformed the other models on the C2 challenge with an average improvement of 9.98\% (\F) and 27.08\% (MAP) over the vanilla BERT, while its dual variant reported gain of only 8.82\% (\F) and 18.49\% (MAP). Likewise for the C3 challenge, vanilla sciBERT outperformed all other models with an average improvement of 1.21\% (\F) and 9.51\% (MAP) over the vanilla BERT architecture, while its dual variant underperformed.
These findings indicate that single-BERT architectures solve RSA related challenges better than dual architecture. Furthermore, the single-BERT architecture is more memory-efficient and trains faster.

\noindent\textbf{RQ3: Does Multitask Learning (MTL) return more accurate results than single-task learning?} 
To answer this RQ we compare results from using single-task versus multi-task learning. Results are reported in Table \ref{icse:RQ3} and show that while  multitask learning generalizes across all datasets better for the C1 and C2 challenge, it does not work well with the C3 challenge. Though vanilla sciBERT outperformed all other variants with an average improvement of 17.72\% (\F) and 20.11\% (MAP) over the vanilla BERT on the C1 challenge, the multitasking variants of BERT and seBERT outperformed their vanilla counterparts. In the C2 challenge, sciBERT\_multi achieved the best performance with an average improvement of 12.95\% (\F) and 30.87\% (MAP) over the vanilla BERT. Moreover, BERT\_multi and seBERT\_multi also outperformed their respective vanilla counterparts.

However, in the C3 challenge, the multitask variants did not improve generalization as compared to their single-task counterparts. We attribute this behavior to the complexity of the task as C3 is a mixture of both the corrupt and missing links. 
One potential problem is that samples which are missing a design element are labeled as incomplete even though all the included links are correct, and this could create negative transference, and inhibit optimization of the overall loss. Negative transference occurs when prior knowledge interferes with new learning \cite{wiki:Negative_transfer_(memory)}. In our case, while we hypothesized that the auxiliary task of link prediction would be helpful, results suggest that it may actually interfere with the RSA learning process in the C3 challenge, resulting in poor optimization.

In summary, we conclude that multitask learning helps generalization in a scenario where the set of design elements are either corrupt or incomplete; however, in datasets with co-mingled problems, negative transfer negatively impacts performance.

\noindent\textbf{RQ4: Which BERT-based model returns the best results for the general RSA challenge?}\\
Finally, based on observations from RQ1, we conclude that the pretrained language models significantly outperform the IR-based approach in solving the RSA challenge. RQ1 showed that sciBERT was the best vanilla LM. RQ2 results showed that feeding additional information to the models using the dual-BERT encoders was not effective as the vanilla BERT outperformed the dual variants, with some exceptions for the C2 challenge. Finally, in RQ3 we experimented with MTL, and found that the multitask variants while outperforming the vanilla BERT on C1 and C2 challenge, suffered from negative transfer on one of the most important challenges i.e. C3. Therefore, these findings suggest that vanilla sciBERT is the best model for automated RSA.

%% file: 07-RelatedWork.tex
\section{Related Work}
\label{sec:RelatedWork}
The architectures behind the RSA models that we have proposed in this study have been previously used in tackling several text-to-text challenges. Lu et al. \cite{lu2020twinbert} proposed TwinBERT, where they used dual encoders to encode the query and documents separately. Similarly, in this study we propose DSat-BERT which closely resembles the architecture of TwinBERT. However, we utilize one of the two encoders to provide  extra information to the model. Caruana et al. \cite{caruana1993multitask, caruana1997multitask} introduced multitask learning with a notion that it might be easier to learn several hard tasks at once, rather than independently. Specifically, MTL often improves generalization by leveraging the domain-specific information contained in the training signals of related tasks \cite{caruana1997multitask}. Motivated by these studies, we introduce MSat-BERT, which learns two tasks i.e. satisfaction prediction and link prediction in parallel while using a shared representation. In this study, we utilize one of the most common approaches i.e. hard-parameter sharing in achieving the shared representation; however, no prior study has compared the single, dual and multitask variants in the field of Software Engineering.

In addressing the RSA challenge, Holbrook et al. \cite{holbrook2006assessing, holbrook2009toward} utilized an IR based approach which first parsed a requirement and all of its linked design elements into chunks, and then calculated the pairwise similarity between them\cite{hacioglu2004semantic}. They  used a domain-specific thesaurus to tag words within each chunk with their associated synonyms, in order to match semantically related concepts, and then evaluated two different techniques for matching design chunks to requirements. These were N\"aive Satisfaction Assessment, and Satisfaction Assessment using TF-IDF. 
While the n\"aive approach relied only on the textual similarity between the chunks, the TF-IDF approach took the importance of a term within a chunk into consideration, and calculated the similarity using the cosine between two vectors. The authors performed experiments with numerous similarity thresholds as a cutoff to categorize the requirement-design pairs as the candidate satisfaction mapping. The candidate pairs were then vetted by the human analyst as part of a final satisfaction assessment mapping, and compared against a golden-answer set.   They extended their work \cite{holbrook2013study} with part-of-speech tagging, and proposed rule-based techniques for matching design chunks to requirements chunks, and implemented their techniques in the Requirements SATisfaction Tool (RESAT).  However, rule-based approaches are brittle with respect to the way artifact text is structured, require manually curated, and evolvable grammar rules, and return limited accuracy. In contrast, the BERT-based approaches presented and evaluated in this paper show significant promise in moving towards and automated RSA solution.

%% file: 08-ThreatstoValidity.tex
\section{Threats TO Validity}
\label{sec:Threats}

There are several threats to validity in this study. First, we only evaluated RSA on five different datasets taken from four different projects and domains. While we cannot draw generalized conclusions based on five datasets, our work represents significant improvements over existing baselines which returned far less accurate results and were evaluated on only two very small datasets (i.e., CM1 and GANTT -- which only has 17 links). As with all studies of this nature, we made several design decisions about specific architectures and language models to include in our experiments. However, these decisions were motivated by results from related studies, such as using BERT approaches to support trace link generation, and represented diverse options. 

Finally, our results suggested the single-BERT architecture delivered the best results; however, performing hyper-parameter tuning could certainly help achieve better results for the dual-BERT architectures, and multitasking variants, but hardware constraints limited our ability to conduct hyper-parameter tuning. Finally, as our goal is to find the best model in terms of both accuracy and efficiency while minimizing resource consumption, we used the same set of parameters for all the models. Future work is likely to build on these results and achieve better results.

%% file: 09-ConclusionandFutureWork.tex
\section{Conclusion and Future Work}
\label{sec:Conclusion}

This study has explored the use of PLMs using several different techniques to automate the process of RSA. Our experimental results showed that their use can significantly outperform the IR-based approach. More specifically, sciBERT outperformed the other general purpose models due to its richness in multi-domain knowledge. In addition, given the same set of hyper-parameters, we concluded that the single-BERT architecture can be quite effective for automated RSA, while the dual-BERT architecture is neither as accurate nor as efficient. Our results also showed that MTL performs better as compared to the vanilla BERT architecture on C1 and C2 challenge. However, it suffers from negative transfer when these challenges coexist and so is often the case. Though all of these models could be used interchangeably to achieve the maximum accuracy for a particular task, setup of a model is an expensive process and requires non-trivial resources. Therefore, from the findings in this study, we conclude that the single BERT architecture (scibert) on average across all the dataset is the best overall model in performing RSA. 

In this paper we focused on quantitative rather than qualitative analysis. However, in future work we will also address the challenge of generating explanations for the RSA decision, and will conduct user studies with software project stakeholders.